\documentclass[preprint,showpacs,preprintnumbers,amsmath,amssymb]{revtex4}
\usepackage{graphicx}
\usepackage{dcolumn}
\usepackage{bm}

\begin{document}
\pagenumbering{arabic}
\preprint{Preprint}
\title{Direct proton decay of the isoscalar giant dipole resonance in $^{208}$Pb }
\author{B. K. Nayak}
\altaffiliation{Permanent address: Bhabha Atomic Research Centre,
Mumbai 400 085, India }
\author{U. Garg}
\author{M. Koss}
\author{T. Li}
\affiliation{%
Department of Physics, University of Notre Dame, Notre Dame, IN 46556, USA }%
\author{E. Martis}
\altaffiliation{%
REU student from Beloit College, Beloit, WI 53511, USA }%
\affiliation{%
Department of Physics, University of Notre Dame, Notre Dame, IN 46556, USA }%
\author{ H. Fujimura}
\affiliation{%
 Research Center for Nuclear Physics, Osaka University, Mihogaoka 10-1,
Ibaraki, Osaka 567-0047, Japan}%
\author{ M. Fujiwara}
\altaffiliation{ Also at Kansai Photon Science Institute, Japan Atomic Energy Agency, Kizu, Kyoto 619-0215, Japan }%
\author{ K. Hara}
\author{ K. Kawase}
\author{K. Nakanishi}
\author{E. Obayashi}
\author{H. P. Yoshida}
\affiliation{%
 Research Center for Nuclear Physics, Osaka University, Mihogaoka 10-1,
Ibaraki, Osaka 567-0047, Japan}%
\author{ M. Itoh}
\altaffiliation{Current address: RI Center, Tohuku University, Sendai 980-8578, Japan}%
\affiliation{%
Department of Physics, Kyoto University, Kyoto 606-8502, Japan}%
\author{ S. Kishi}
\affiliation{%
Department of Physics, Kyoto University, Kyoto 606-8502, Japan}%
\author{ H. Sakaguchi }
\affiliation{%
Department of Physics, Kyoto University, Kyoto 606-8502, Japan}%
\author{ H. Takeda}
\affiliation{%
Department of Physics, Kyoto University, Kyoto 606-8502, Japan}%
\author{ M. Uchida}
\affiliation{%
Department of Physics, Kyoto University, Kyoto 606-8502, Japan}%
\author{ Y. Yasuda}
\affiliation{%
Department of Physics, Kyoto University, Kyoto 606-8502, Japan}%
\author{M. Yosoi}
\affiliation{%
Department of Physics, Kyoto University, Kyoto 606-8502, Japan}%
\author{ R. G. T. Zegers}
\affiliation{National Superconducting Cyclotron Laboratory and Department of Physics and Astronomy, Michigan State University, East Lansing,  MI 48824, USA}
\author{ H. Akimune }
\affiliation{%
Department of Physics, Konan University, Kobe 658-8501, Japan}%
\author{ M. N. Harakeh }
\affiliation{%
Kernfysisch Versneller Instituut, University of Groningen, 9747 AA Groningen, The Netherlands }%
\author{ M. Hunyadi}
\altaffiliation{Permanent address: Institute of Nuclear Research of the Hungarian Academy of Sciences, H-4001 Debrecen, Hungary}%
\affiliation{%
Kernfysisch Versneller Instituut, University of Groningen, 9747 AA Groningen, The Netherlands }%
\date{\today}

\begin{abstract}

The excitation and subsequent proton decay of the isoscalar giant dipole
 resonance (ISGDR) in $^{208}$Pb have been investigated via the
$^{208}$Pb($\alpha, \alpha^{\prime}p)^{207}$Tl
 reaction at 400 MeV. Excitation of the ISGDR has been identified by the difference-of-spectra method. The enhancement of the ISGDR strength at high excitation energies observed
 in the multipole-decomposition-analysis of the singles $^{208}$Pb($\alpha, \alpha^{\prime}$) 
spectra is not present in the excitation energy spectrum obtained in coincidence measurement. The partial branching
 ratios for direct proton decay of ISGDR to low-lying states of $^{207}$Tl 
have been determined and the results are compared with predictions of 
continuum random-phase-approximation (CRPA) calculations.

\end{abstract}


\maketitle

The compression-mode isoscalar giant dipole resonance (ISGDR), like the isoscalar 
giant monopole resonance (ISGMR), provides a direct method to obtain the incompressibility
of the nucleus and of nuclear matter (K$_{\rm{nm}}$)\cite{1}. The first direct evidence for the ISGDR in
nuclei was obtained in $^{208}$Pb with the ``difference-of-spectra'' technique (DOS) in inelastic scattering of 200 MeV $\alpha$-particles \cite{2}. Subsequently, the Texas
A $\&$ M group obtained the ISGDR strength distributions in several nuclei using a
multipole-decomposition
 analysis (MDA) of the background-subtracted inelastic $\alpha$-scattering
spectra at 240 MeV \cite{3,4,5}.
 However, a major concern was that the centroids of the ISGDR strength distribution from these
 studies were found to be consistently lower than those predicted by calculations employing the same value of nuclear
 incompressibility that correctly reproduced the centroid energies of the ISGMR strength
 distributions. This ambiguity has been resolved
 by more precise and instrumental-background-free measurements of ISGDR strength
distributions using inelastic scattering of 400 MeV $\alpha$
particles \cite{6,7,8,9,10,11}. The value of the nuclear incompressibility,
K$_{\rm{nm}}$, obtained from the ISGDR data is now fully consistent with that
from the ISGMR data. 
 However, a problem has remained with the ISGDR strength extracted from the aforementioned singles measurements: significantly large E1 strength
 was observed at the higher-excitation energy part of the main ISGDR peak in all
nuclei under investigation (see, for example, Fig.~3 of Ref.~\cite{10} and Fig.~4 of Ref.~\cite{11}). This extra strength was
attributed to contributions to the continuum from three-body channels, such as
pick-up/breakup reactions. These processes form part of the continuum and
lead to spurious ISGDR contributions in the MDA
because of the forward-peaked nature of their angular distributions \cite{8,10,11}. The effect is
significant in the high excitation-energy region where the associated
cross sections are very small. Incidentally, a similar increase at higher
excitation energies has been reported in the E0 strength in $^{12}$C as well
when a MDA was carried out without subtracting the continuum from
the excitation-energy spectra \cite{12}.

A primary motivation of the present study has been to address the
aforementioned problem of excess ISGDR strength at high excitation energies 
apparent in the singles measurements. Investigations of the direct
 proton-decay channels of ISGDR in the ($\alpha, \alpha^{\prime}p$)
reaction renders this eminently possible. In general, decay studies afford several
advantages over the singles measurements. Firstly, events in the continuum that are associated with 
forward-peaking processes, such as quasifree scattering and pick-up/breakup
reactions, are effectively suppressed by putting a coincidence condition with
decay protons at backward angles; the contributions from the resonance, however,
remain in the true coincidence spectra (subject, of course, to the decay
probabilities). Secondly, by gating on the particle decay channels populating
specific single-hole states in the daughter nuclei, the resonance states, which 
are comprised of states with 1p-1h configurations, are enhanced, providing a
detailed look at the decay properties, such as the relative population and
strength of
particle-decay channels from the resonance to the final hole states in the
daughter nucleus. This can provide crucial information to test the available
microscopic model calculations.

In the past, direct proton-decay from the ISGDR in $^{208}$Pb has been measured using
($\alpha, \alpha^{\prime}p$) reaction at a bombarding energy of 200 MeV \cite{13,hun2}. In these
measurements, a new  $L=$2 resonance 
was reported at E$_{\rm{x}}$=26.9$\pm$0.7 MeV with a width of 6.0$\pm$1.3 MeV; this has been suggested as the $L=2$ compression-mode resonance \cite{14}.
In a subsequent neutron-decay measurement at the same energy, although 
the population of the
low-lying states through direct neutron decay of the ISGDR region was found to
be significant and well separated from the statistical decay, the direct neutron decay from the continuum region
above the ISGDR energy did not show any significant population to the final
single-hole states \cite{15}. The present investigation not only provides
further information on the decay of the ISGDR to specific particle-hole states,
but also allows another look at the aforementioned $L=$2 resonance which had
not been identified in any of the singles measurements.

 Angular distributions of differential cross sections of ($\alpha, \alpha^{\prime}$) reactions strongly depend on the transferred angular momentum $L$. This is evident in Fig.~1, where the differential cross sections for various multipoles ($L \le$ 3) for the $^{208}$Pb($\alpha, \alpha^{\prime}$) reaction
 at E$_{\rm{\alpha}}$= 400 MeV calculated in DWBA in the angular
 range 0$^{\circ}$--3$^{\circ}$ for excitation energy E$_{\rm{x}}$=22.5 MeV are
seen to be clearly distinct from each other. The cross section for ISGDR ($L$=1)
is maximal at 1.28$^{\circ}$; the cross sections for the $L$=2 and $L$=3 modes, 
on the other hand, are more or less constant over 0$^{\circ}$--3$^{\circ}$. By
subtracting the differential cross section near 0$^{\circ}$ from that at small
finite angles (1$^{\circ}$-1.5$^{\circ}$), one can identify the ISGDR strength
since the contributions from $L$=2 and $L$=3 modes would be essentially
eliminated in the subtraction process. This forms the basis of the DOS
technique which has been successfully employed in the past to identify GMR and 
ISGDR strengths in small-angle inelastic scattering spectra \cite{2}. 

 Guided by these calculations for the cross sections, measurements were performed at the
 ring cyclotron facility of Research Center for Nuclear Physics (RCNP), at Osaka University, which provided an
 $\alpha$-particle beam with an energy of 400 MeV. The $\alpha$-beam bombarded a $^{208}$Pb target with a 
thickness of 10.1 mg/cm$^{2}$. The ejectiles were momentum analyzed by the magnetic spectrometer
 Grand Raiden \cite{16} which was set at 0$^{\circ}$ to the beam direction covering a scattering angular
 range 0$^{\circ}$ to 1.8$^{\circ}$; the total solid angle in this setting was 1.2 msr. The
``energy bite'' of the spectrometer at this energy is ~30 MeV; data were, therefore,
obtained over the excitation-energy range $\sim$4--$\sim$34 MeV. Decay protons were
detected in coincidence with inelastically scattered $\alpha$ particles with
sixteen silicon solid-state detectors, each with a thickness of 5.0 mm, placed
at angles of 112.5$^{\circ}$, 135$^{\circ}$
and 157.5$^{\circ}$ with respect to the beam direction. The energy calibrations of the singles ($\alpha, \alpha^{\prime}$) spectra and of decay-proton energies measured with solid-state detectors were carried out by using the $^{12}$C($\alpha, \alpha^{\prime}$) and
 $^{12}$C($\alpha, \alpha^{\prime}p$) reactions.
 
\begin{figure}
\includegraphics[width=8.0cm]{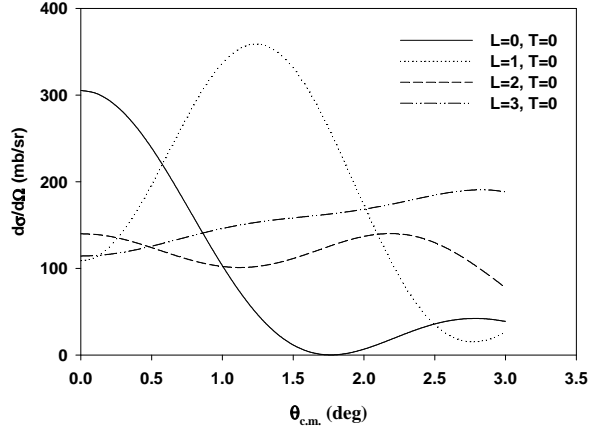}
\caption{Differential cross sections for exciting isoscalar giant resonances via the $^{208}$Pb($\alpha, \alpha^{\prime}$)
reaction calculated in DWBA, assuming 100$\%$ exhaustion of the energy-weighted sum rules for the respective transitions.
 The calculations were performed at E$_{\alpha}$ = 400 MeV for excitation energy E$_{\rm{x}}$=22.5 MeV.
 The curves drawn are for ISGMR (solid), ISGDR (dotted), ISGQR (dashed), and HEOR (dash-double-dotted).}
\label{fig1}
\end{figure}

\begin{figure}
\includegraphics[]{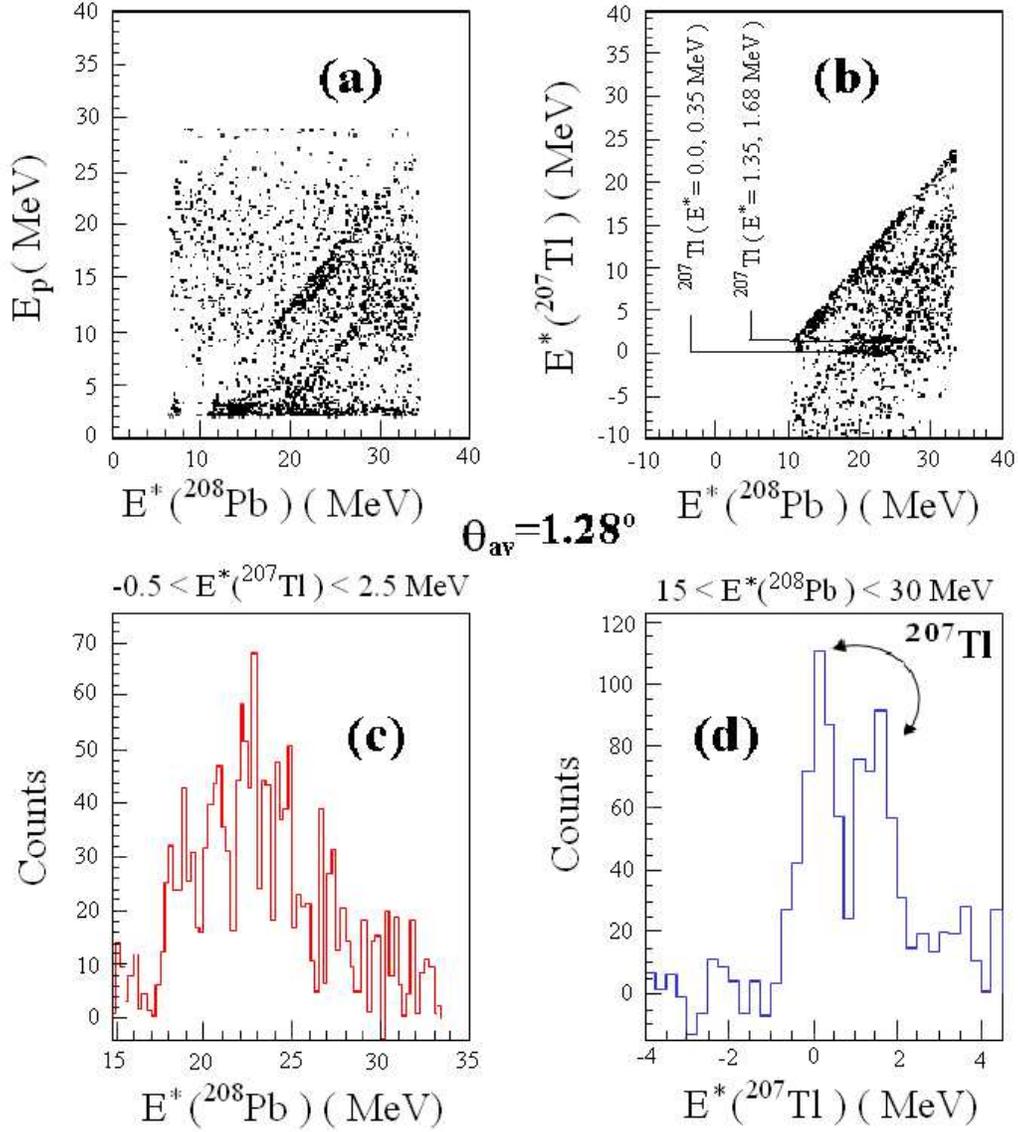}
\caption{(a) Two-dimensional scatter plot of the decay proton energy E$_{\rm{p}}$ versus excitation
 energy E$^{\rm{*}}$ in $^{208}$Pb. (b) Same as (a), but the proton energies have been converted
 to excitation energies in $^{207}$Tl. The loci correspond to decay-events to specific
 final states in $^{207}$Tl. (c) The projection of data in Fig.~2 (b) onto the excitation-energy axis in $^{208}$Pb for the indicated
 excitation-energy range in $^{207}$Tl. The spectrum shows primarily the ISGDR
 strength distribution. (d) The projection of data in Fig.~2 (b) onto the final-state excitation-energy axis in $^{207}$Tl. The first peak is a composite of the ground state and the 0.35 MeV
 excited state; the second peak is a combination of the 1.35 MeV and 1.67 MeV excited states.}
\label{fig2}
\end{figure}

Fig.~2(a) shows the two-dimensional scatter plot of decay-proton energy versus the target
excitation
 energy. One can clearly see several loci corresponding to the proton-decay channels populating
the low-lying final states in $^{207}$Tl. Fig.~2(b) is essentially the same as
Fig.~2(a) but shows the correlation between the energies of the final
 states in $^{207}$Tl and the target excitation energy in $^{208}$Pb instead; Fig.~2(c) 
shows the excitation-energy spectrum of $^{208}$Pb in coincidence with proton decay to low-lying states in $^{207}$Tl. Fig.~2(d)
 shows the final-state spectrum of $^{207}$Tl for proton decay. 

 In Fig.~3, the final-state spectrum has been decomposed to show the relative population of
 various low-lying states. The FWHM of individual states in the final-state excitation-energy spectra for $^{207}$Tl was $\sim$650 keV, as obtained from multiple-peak fitting of the excitation-energy spectrum;  it was not possible, hence, to clearly resolve clearly the individual final states in $^{207}$Tl.  However, there are two enhanced bumps,
 denoted as groups A and B, attributed to proton decays to the
( 3s$_{\rm{1/2}}$+2d$_{\rm{3/2}}$) hole states and to the (1h$_{\rm{11/2}}$+2d$_{\rm{5/2}}$) hole states, respectively. In addition,
 a weak bump structure is observed at the location of the 1g$_{\rm{7/2}}$ hole state. 
 
\begin{figure}
\includegraphics[]{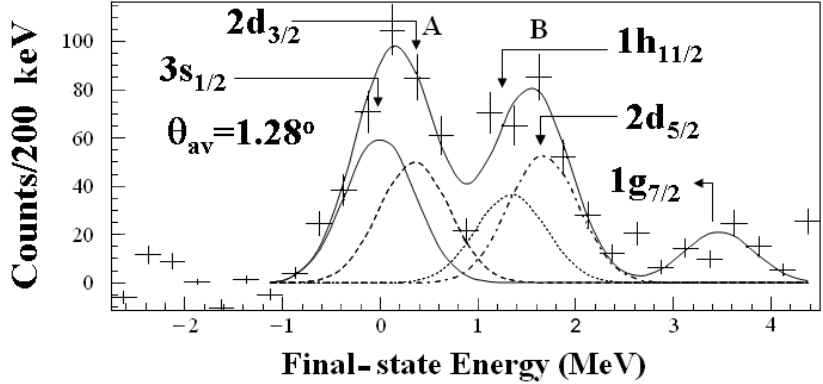}
\caption{The final-state spectrum of the low-lying proton-hole states in $^{207}$Tl generated
 by gating on the ($\alpha, \alpha^{\prime}$) events in the ISGDR excitation-energy region at the
 scattering angles $\theta_{\rm{c.m.}}$= 1.0$^{\circ}$ -1.5$^{\circ}$. }
\label{fig3}
\end{figure}

The  proton-decay branching ratios to final states can be obtained
from the ratio of the energy-integrated  double-differential cross sections for 
coincidence with decay protons to final states  and the singles cross section in the
excitation energy range of 15.0--30.0 MeV. Such an analysis has been carried out for the
three groups of final states of $^{207}$Tl corresponding to the
(3s$_{\rm{1/2}}$+2d$_{\rm{3/2}}$), (1h$_{\rm{11/2}}$+2d$_{\rm{5/2}}$), and 1g$_{7/2}$, as
shown in Fig.~3. The singles double-differential cross sections were obtained from
previous experimental results \cite{7}. The partial branching ratios for proton decay
obtained for these final states are listed in Table~\ref{comparison}. The errors quoted in
the values of these branching ratios correspond only to the statistical uncertainties in
the singles and coincidence double differential cross sections. Since the singles and
coincidence measurements were performed with the same set-up and basically under identical 
conditions, the systematic errors would, in principle, 
cancel out in the ratios or, at any rate, be reduced greatly compared to the statistical errors.
The total experimental
proton-decay branching ratio to the low-lying states of $^{207}$Tl is found to be
(1.97$\pm$0.3)$\%$. The predictions of recent CRPA calculations for the partial branching
ratios, b$_{\rm{i}}$,  for proton decay of the ISGDR in the excitation energy range
15.0--35.0 MeV in $^{208}$Pb to the 3s$_{\rm{1/2}}$, 2d$_{\rm{3/2}}$, 1h$_{\rm{11/2}}$,
2d$_{\rm{5/2}}$ and 1g$_{\rm{7/2}}$ hole states in $^{207}$Tl are
also given in Table~\ref{comparison}, with a theoretical value totaling 2.51$\%$
\cite{17}. 
For comparison, the branching ratios from Ref.~\cite{hun2} for
direct proton decay of the ISGDR in the excitation energy range 19.0--25.0 MeV in
$^{208}$Pb are provided as well. 

\newcommand\T{\rule{0pt}{2.6ex}}
\newcommand\B{\rule[-1.2ex]{0pt}{0pt}}
\begin{table}[h]
\caption{The experimental value of partial  branching ratios (B. R.) for direct proton-decay from
the ISGDR in the excitation-energy range of 15.5 MeV to 30.5 MeV in $^{208}$Pb to the various
proton-hole states in $^{207}$Tl are given in percent. The theoretical branching ratios
b$_{\rm{i}}$ \cite{17}, which refer to the theoretical strength
of ISGDR in the excitation-energy range of 15.0 MeV to 35.0 MeV, are also given. The values of
branching ratios from previous work~\cite{hun2}
are provided for comparison.}
\ \\
\begin{tabular}{ccccccccccc}
\hline
\;\;\;\;Final state \T \B \;\;\;\;& \;\;\;E$_{fs}$\;\;\;\;&$\;\;\;\;\ B.R._{\rm{exp}}\footnote[1]{This work} \;\;\;\;\;\;$&\;\;\;\;\;b$_{\rm{i}}\footnote[2]{ Ref.~\cite{17}} $\;\;\;\;&\;\;\;\;\;\;$B.R.\footnote[3]{Ref.~\cite{hun2} (see text)}\;$\;\;\;\;\\
\;\;\;\;\T \B \;\;\;\;& \;\;\;(MeV)\;\;\;\;&$\;\;\;\;\; \;\;\;\;\;\;$&\;\;\;\;\;\;\;\;\;&\;\;\;\;\;\; \;\;\;\;\;\\
\hline
3s$_{\rm{1/2}}$ \T \B& 0.0 & & 0.65$\%$ \\
$$\T\B& &0.74$\pm$0.15$\%$& &2.3$\pm1.1\%$ \\
2d$_{\rm{3/2}}$ \T \B& 0.35 & & 0.80 $\%$\\
\hline
1h$_{\rm{11/2}}$ \T \B& 1.35 & & 0.29$\%$\\
 $$\T\B& &0.99$\pm$0.20$\%$& &$1.2\pm0.7\%$ \\
2d$_{\rm{5/2}}$ \T \B& 1.67 & & 0.75$\%$ \\
\hline
1g$_{\rm{7/2}}$ \T \B& 3.47 &$0.24\pm0.10\%$ &0.02$\%$ \\
\hline
\end{tabular}
\label{comparison}
\end{table}

\noindent

In this work, we have employed the experimental singles differential
cross section obtained from Ref.~\cite{7} for extracting the branching ratios. In
Ref.~\cite{13}, on the other hand, the branching ratios could not be deduced directly
because the corresponding singles cross sections for the ISGDR were not available.
Instead, the partial differential cross sections for proton decay were
converted directly to sum-rule strengths by comparison with the calculated DWBA cross
sections for 100$\%$ EWSR of the ISGDR located at the relevant excitation energy;
these sum-rule strengths were, then, stated as the branching ratios. This, of course, was subject to the uncertainties associated with DWBA calculations, including also the uncertainty in the centroid energy of the ISGDR.  In a later analysis \cite{hun2}, the situation was rectified by comparing the proton decay to specific states to the total decay cross section, including all direct proton and neutron decays, and the statistical decays. The branching ratios included in Table~\ref{comparison} are from Ref. \cite{hun2} and are, essentially, equivalent to the true branching ratios obtained in the present work.
We find that while the branching ratios from the two measurements are in satisfactory agreement for the
(1h$_{\rm{11/2}}$+2d$_{\rm{5/2}}$) hole states, there remains a significant
discrepancy
for the (3s$_{\rm{1/2}}$+2d$_{\rm{3/2}}$) states. Also, the observed branching ratios are in reasonable
agreement with the CRPA predictions \cite{17} except for the 1g$_{7/2}$ state, where the
experimental branching ratio is larger by an order of magnitude.
 
 Fig.~4(a) shows the double-differential cross section spectrum for
$\theta_{\rm{av}}$=1.28$^{\circ}$, corresponding to the maximum of ISGDR cross
section, and Fig.~4(b) for the forward angular range
0.0$^{\circ}$--0.3$^{\circ}$ ($\theta_{\rm{av}}$=0.19$^{\circ}$). The
subtraction spectrum is shown in Fig.~4(c) after taking into account the
excitation energy dependence of the transmission probabilities. While the
statistics in the subtraction spectrum are rather weak, it clearly shows the
contribution from the ISGDR component, as indicated by the solid line in
Fig.~4(c), which is a Gaussian fit to the data. The centroid energy of the ISGDR
peak in the subtraction spectrum is 22.7$\pm$1.6 MeV which is in 
agreement with the value obtained in the singles measurement (22.7$\pm$0.2 MeV)~\cite{7}. 
It may be noted that, within the limited statistics of this
experiment, there also are bumps in the excitation-energy spectrum for
$\theta_{\rm{av}}$=0.19$^{\circ}$ both above and below the ISGDR region
(see Fig.~4(b)). These bumps have been  identified in the previous proton-decay
measurements \cite{13,hun2} as corresponding, respectively, to $L=2$ (the
higher-lying bump) and $L=3$ excitation (the lower-lying bump). As mentioned
earlier, and shown in Fig.~1, the $L$=1 excitation cross section is
significantly larger when compared with the $L$=2 excitation cross section in the angular
range 1.0$^{\circ}$ to 1.5$^{\circ}$ covered in our measurement. This might be a reason
why the $L$=2 strength above ISGDR region reported in the previous particle-decay
measurement \cite{13,hun2,14} is not seen as
prominently in our data. Still, to get a quantitative estimate of the $L$=2 contribution,
the excitation energy spectrum in the range of 15.5--33.5 MeV was fitted with three
Gaussian distributions. A free fit resulted in peaks with centroid energies of 17.2 MeV,
 22.7 MeV, and 29.6 MeV, respectively, with the width of each peak fixed at 3.8 MeV. That 
would put the ``$L$=2'' peak in our data at an energy slightly higher than that reported in
Ref.~\cite{13}. It can be stated nonetheless that there exists some strength in our spectra
at the location of the previously-reported $L$=2 peak. We have also carried out a
three-peak fit by keeping the centroid energy of the middle peak
fixed at 22.7 MeV and the other two centroid energies fixed at the values reported for 
the $L$=2 and $L$=3 peaks in Refs.~\cite{13,hun2}; the resulting fit is shown in Fig.~4(a). The
ratio of the integrated areas of the ``$L$=2'' Gaussian to the sum of the
areas of the three Gaussian peaks suggests that the contribution of the $L$=2 multipole
in the aforementioned excitation-energy range is around 22$\%$, not inconsistent with the 
results of Refs.~\cite{13,hun2}, once the likely angular-distribution effects are taken into
account. 

\begin{figure}
\includegraphics[scale=0.6]{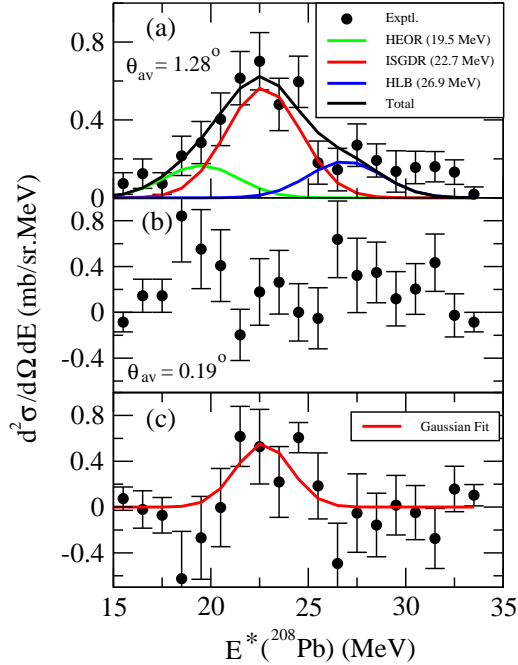}
\caption{(a) The double-differential cross section spectrum for the scattering angular range
$\theta_{\rm{c.m.}}$= 1.0$^{\circ}$ -1.5$^{\circ}$ and in the excitation-energy
range  -0.5 MeV $<$ $^{207}$Tl $<$ 2.5 MeV. The solid lines are 3-Gaussian fits
to the data (see text).
 (b) Same as (a) but for the scattering angular range $\theta_{\rm{c.m.}}$= 0.0$^{\circ}$-0.3$^{\circ}$.
 (c) The subtraction spectrum of (a) and (b). The subtraction spectrum 
clearly shows the contribution from the ISGDR component. }
\label{fig4}
\end{figure}
 
In summary, the excitation and proton-decay of the ISGDR has been measured using the
$^{208}$Pb($\alpha, \alpha^{\prime}p$)$^{207}$Tl reaction at 400 MeV. The large ISGDR
strength observed at the highest excitation energies that was observed in the singles
measurements is not present in the coincidence spectra, confirming the
suggestion then advanced that this excess strength was spurious and arose from other, non-resonant,
phenomena. The ISGDR centroid energy of 22.7$\pm$1.6 MeV, obtained from the coincidence
measurements, is in agreement with the value of 22.7$\pm$0.2 MeV from the singles
measurements. The total proton-decay branching ratio to low-lying states of $^{207}$Tl is
found to be (1.97$\pm$0.3)$\%$, in reasonable agreement with the CRPA prediction of
2.51$\%$ for this value.

The authors acknowledge the staff of the RCNP ring cyclotron for providing
 a high-quality, halo-free $\alpha$ beam during the experiment. This work was supported
in part by the US-Japan Cooperative Science Program of the JSPS, the US National
Science Foundation (grant Nos. INT-9910015, PHY04-57120 and PHY07-581100),
and the University of Notre Dame.


\end{document}